\documentstyle[12pt]{article}

\author{M.A.Semenov-Tian-Shansky\thanks{The present paper is a
translation of an article originally published in Zapiski Nauchn.
Semin. Pomi, vol. 200, SPb. 1993. }\\
Physique Math\'{e}matique, \\
Universit\'{e} de Bourgogne 21004 Dijon France \\
and St. Petersburg Branch\\
of the V.A.Steklov Mathematical Institute \\
St.Petersburg 191011 Russia}
\title{Monodromy Map and Classical r-matrices}
\newtheorem{theorem}{Theorem}
\newtheorem{proposition}{Proposition}
\newtheorem{lemma}{Lemma}
\newtheorem{definition}{Definition}

\begin{document}

\maketitle
\begin{abstract}
We compute the Poisson bracket relations for the monodromy matrix of the
auxiliary linear problem. If the basic Poisson brackets of the model
contain derivatives, this computation leads to a peculiar kind of symmetry
breaking which accounts for a 'spontaneous quantization' of the
underlying global gauge group. A classification of possible patterns of
symmetry breaking is outlined.
\end{abstract}

\section{Introduction}

Computation of the Poisson bracket relations for the monodromy matrix of
the auxiliary linear problem is one of the keypoints in the Classical
Inverse  Scattering Method \cite{FadTak}. Its quantum counterpart, the
computation of the commutation relations for the quantum monodromy
matrix, plays an equally  important role in Quantum Inverse Scattering
Method. Usually, this computation is carried out under the technical
assumption of 'ultralocality'of the basic Poisson brackets (a precise
definition is given below in Section 1, Lemma 2.)

It is the aim of the present note to discuss the general case. As we shall
see, dropping out of the ultralocality condition leads to new physical
effects (some kind of spontaneous symmetry breaking). In quantum setting
these effects manifest themselves in spontaneous quantization (i.e.
deformation of special kind) of certain symmetry groups. Similar effects
were discovered recently in Conformal Field Theory where they have drawn
much attention \cite{Alv}, \cite{Bab}, \cite{Gaw}. We shall see that
possible patterns of symmetry breaking may be fully classified: an easy
reformulation reduces the problem to the classification of solutions of
the modified classical Yang--Baxter equation for the square of the
original Lie algebra and hence to the classification theorem of Belavin
and Drinfeld \cite{BelDr}.

\section{Monodromy Matrices and Zero Curvature Equations}

Zero curvature equations are compatibility conditions for the linear
system of differential equations
\begin{equation}
\label{flat}
\begin{array}{c}
\partial _x\psi =L\psi , \\
\partial _t\psi =A\psi .
\end{array}
\end{equation}

In applications to integrable systems the coefficients $(L,A)$ of the flat
connection usually take values in the loop algebra
{\bf g}$={\cal L({\bf a})}$
 of an auxiliary finite-dimensional Lie semi-simple algebra
 ${\bf a.\ }$ The main content of the Inverse Scattering Method as
 applied to zero curvature   equations is given by the following two
 assertions:

(1){\em Integrals of motion for the zero curvature equations are spectral
invariants of the ordinary differential operator $\partial _x-L.$}

(2){\em These integrals are in involution with respect to some natural
Poisson bracket on the phase space. }(The notion of 'natural' Poisson
Brackets will be precised below.)

Let us suppose for concreteness that the coefficients of $L$ are periodic
functions of $x$ with period $2\pi .$ In that case, according to the
classical Floquet Theorem, spectral invariants of $L$ depend only on the
eigenvalues of the monodromy matrix. By definition, the monodromy
matrix of the differential equation
\begin{equation}
\label{L}\partial _x\psi =L\psi
\end{equation}

is given by
\begin{equation}
\label{Mon}M=\psi (2\pi )\psi (0)^{-1},
\end{equation}

where $\psi $ is a fundamental solution of (2). It is natural to
regard $M$ as an element of the Lie group $G$ which corresponds to
${\bf g.\ }$Let us consider the {\em monodromy map }
$M:{\cal M\longrightarrow }G$ which assigns to a point $L\in {\cal M}$
of the phase space the corresponding monodromy matrix. Since the
eigenvalues of $M$ are integrals of motion, the monodromy
itself satisfies an evolution equation of the form%
$$
\partial _tM=\left[ M,N\right]
$$

which is usually called the Novikov equation.

Let now{\cal \ }${\cal F_{{\it t}}}$ $:{\cal M\longrightarrow M}$ be the
dynamical flow associated with the zero curvature equation, and $%
F_t:G\longrightarrow G$ the flow on $G$ determined by the Novikov equation.
The relation between the two flows is given by the commutative diagram
\begin{equation}
\label{diag}
\begin{array}{ccccc}
&  & {\cal F_{{\it t}}} &  &  \\
& {\cal M} & \longrightarrow & {\cal M} &  \\
M & \downarrow & F_t & \downarrow & M \\
& G & \longrightarrow & G &
\end{array}
\end{equation}

In other words, the flow ${\cal F_{{\it t}}\ }$factorizes over $G.$

Let us recall now that the phase space ${\cal M}$ is equipped with a
Poisson bracket. It is natural to expect that the group $G$ may also be
equipped with a Poisson structure in such a way that the monodromy map
preserves the Poisson brackets and the diagram (4) consists of Poisson
mappings. In order  to make this picture more precise let us recall first
the hamiltonian interpretation of the spectral invariants of the
monodromy \cite{ReyS}, \cite{S}.

Let ${\bf g}$ be a Lie algebra equipped with a nondegenerate invariant
inner product. Let ${\cal G=}{\it C^\infty (S^1;{\bf g})}$ be the
corresponding current algebra (the algebra of smooth periodic
functions with values in $ {\bf g}$ and with pointwise commutator).
The bilinear form
\begin{equation}
\label{cocyc}\omega (X,Y)=\int \left\langle X,\partial _xY\right\rangle dx
\end{equation}
is a 2-cocycle on ${\cal G}$ and defines a central extension ${\cal G}%
^{\wedge }={\cal G}\uplus {\bf R}$ of ${\cal G\ }$. Using the inner
product on ${\cal G}$ we may identify the dual of ${\cal G}^{\wedge }$
with ${\cal G\oplus }{\bf R}$ ; the coadjoint representation of
${\cal G}$ in ${\cal G}%
^{\wedge }{\cal ^{*}}$ is then given by
\begin{equation}
\label{ad}ad^{*}X(L,e)=(\left[ X,L\right] -e\partial _xX,0).
\end{equation}
The number $e\in {\bf R\ (}$the central charge) is a coadjoint invariant;
without loss of generality we may assume that $e=1$ (i.e. fix an invariant
hyperplane ${\cal \ {\cal G}_{{\it 1}}^{\wedge *}=(G,{\it 1})\subset \ {\cal %
G}^{\wedge *}}$). As usual, the dual of ${\cal \ G}^{\wedge }$ is equipped
with a natural Poisson bracket, the Lie--Poisson bracket of ${\cal \ G}%
^{\wedge }$. Recall that a function is called a Casimir function of a
given Poisson structure if its Poisson brackets with any other function
are identically zero (i.e. if it lies in the center of the Lie algebra of
functions with respect to the Poisson bracket). The Casimir functions
form a
ring with respect to pointwise multiplication.

\begin{theorem}
The ring of Casimir functions on
${\cal \ {\cal G}_{{\it 1}}^{\wedge *}\ }$%
is generated by the spectral invariants of the monodromy matrix of
equation (2).
\end{theorem}

More precisely, let $\varphi \in C^\infty (G)$ be a central function on
 $G$ (i.e. $\varphi (xy)=\varphi (yx)$ for any $x,y\in G$). Then
 $L\longmapsto\varphi (M\left[ L\right] )$ is Casimir function on
 ${\cal \ {\cal G}_{{\it 1 }}^{\wedge *},\ }$ and the ring of Casimir
 functions is generated by functions of this form.

In order to get nontrivial equations of motion from the spectral
invariants
of monodromy we need a different Poisson bracket. The corresponding
construction is basic in the theory of classical r-matrices. Recall that
classical r-matrix on a Lie algebra ${\bf g}$ is a linear operator
$R\in End({\bf g)}$ such that the bracket on ${\bf g}$ given by
\begin{equation}
\label{Rbr}\left[ X,Y\right] _R=\frac 12\left[ RX,Y\right] +\frac 12\left[
X,RY\right]
\end{equation}
satisfies the Jacobi identity. In this case there are two structures of a
Lie algebra on the linear space {\bf g} given by the original Lie bracket
and by the $R-$bracket, respectively. Let
${\cal G_{{\it R}}=}{\it C^\infty(S^1;}{\bf g)}$
be the corresponding current algebra.

\begin{proposition}
\cite{S} The bilinear form on ${\cal G_{{\it R}}}$ given by
\begin{equation}
\label{rcocyc}\omega _R(X,Y)=\frac 12\left( \omega (RX,Y)+\omega
(X,RY)\right)
\end{equation}
is a 2-cocycle on ${\cal G_{{\it R}}.}$
\end{proposition}

Let ${\cal \ {\cal G}_{{\it R}}^{\wedge }}$ be the corresponding central
extension. Clearly, the dual spaces of ${\cal \ G}^{\wedge }$ and ${\cal \
{\cal G}_{{\it R}}^{\wedge }}$ coincide. Hence the space
 ${\cal \ {\cal G}^{\wedge *}\ }$ (and even the hyperplane
 ${\cal \ {\cal G}_{{\it 1}}^{\wedge *}}$) is equipped with two
 different Lie--Poisson brackets which correspond
to the original Lie bracket in ${\cal \ G}^{\wedge }$ and to the $R-$%
bracket, respectively.

\begin{theorem}
The Casimir functions of ${\cal \ G}^{\wedge }$ are in involution with
respect to the $R-$bracket and give rise to zero curvature equations on $%
{\cal \ {\cal G}_{{\it 1}}^{\wedge *}.}$
\end{theorem}

{\em Sketch of a proof.} We shall start with the following Lemma which
 will
be useful in the sequel. Let $\varphi \in C^\infty $ (G). Consider the
functional $\varphi ^M$ on ${\cal \ {\cal G}_{{\it 1}}^{\wedge *}}$
given by
$\varphi ^M:L\longmapsto \varphi (M\left[ L\right] $ ). The Frechet
derivative of $\varphi ^M$ is a function
{\rm grad}$\varphi ^M=$$X_\varphi $
on $\left[ 0,2\pi \right] $ with values in ${\bf g}$ defined by the
relation%
$$
\int \left\langle X_\varphi (x),\xi (x)\right\rangle dx=\left( \frac
d{dt}\right) _{t=0}\varphi \left( M\left[ L+t\xi \right] \right)
$$
for any $\xi \in {\cal G.}$

\begin{lemma}
(i) The Frechet derivative of $\varphi ^M$ is equal to
\begin{equation}
\label{Frechet}X_\varphi =\psi \ \nabla _\varphi \psi ^{-1},
\end{equation}
where $\psi $ is the fundamental solution of (1) normalized so that $\psi
(0)=1,$ and $\nabla _\varphi $ is the left gradient of $\varphi $ on $G$
defined by%
$$
\left\langle \ \nabla _\varphi (x),\eta \right\rangle =\left( \frac
d{dt}\right) _{t=0}\varphi (e^{t\eta }x).
$$
(ii) It satisfies the differential equation
\begin{equation}
\label{eq}\partial _xX_\varphi =\left[ L,X_\varphi \right]
\end{equation}
and the boundary condition
\begin{equation}
\label{bound}X_\varphi (2\pi )=AdL\cdot X_\varphi (0).
\end{equation}
(iii) If $\varphi $ is central on $G$, then X$_\varphi $ is a smooth
function on the circle.
\end{lemma}

The proof of the lemma consists in the standard use of variation of
coefficients in equation (1). Assertions (ii, iii) immediately follow from
the explicit formula (9).

The Poisson bracket of two functionals is a bilinear form of their Frechet
derivatives. The linear operator associated with this bilinear form is
called the Poisson operator. It is easy to write down its explicit
expression for the $R-$bracket in question.

\begin{lemma}
(i) The Poisson operator associated with the $R-$bracket is given by
\begin{equation}
\label{Op}{\cal H_{{\it R}}=(}{\it R+R^{*})\partial _x+(R^{*}\circ
adL+adL\circ R).}
\end{equation}
(ii)\ This operator is bounded if and only if $R=-R^{*}.$
\end{lemma}

The latter case plays an important role; we say then that the Poisson
bracket is {\em ultralocal }and the classical $R-$matrix satisfies the
{\em unitarity condition.}

In the general case in order to get a {\em bona fide }linear operator we
must add to the differential expression (12) some boundary conditions.

\begin{lemma}
Operator ${\cal H_{{\it R}}\ }$is essntially skew-selfadjoint on the space
of smooth periodic functions on $\left[ 0,2\pi \right] .$
\end{lemma}

\begin{definition}
A functional $\Phi $ on ${\cal \ {\cal G}_{{\it 1}}^{\wedge *}}$ is called
smooth if its Frechet derivative lies in the domain of
${\cal H_{{\it R}}.}$
\end{definition}

Lemma 1 (ii) shows that if a function $\varphi \in C^\infty (G)$ is
central,
the corresponding functional $\varphi ^M$ is smooth; hence the Poisson
bracket of such functionals is well defined. Put $X_i=grad\varphi
_i^M,i=1,2. $ To prove that the functionals $\varphi ^M$ are in involution
with each other we have to compute the bilinear form $\left( \ {\cal H}%
_RX_1,X_2\right) .$ By Lemma 1 (ii, iii) the integrand is a total
derivative
of a periodic function.

Assume that $R=-R^{*}$. In this case the Poisson bracket $\left\{ \varphi
_1^M,\varphi _2^M\right\} $ is well defined for any two functions $\varphi
_1,\varphi _2\in C^\infty (G).$ Using Lemma 1 it is easy to check that in
this case $\left( \ {\cal H}_RX_1,X_2\right) $ is still a total
derivative;
however, in this case the boundary terms in general do not vanish. We get
\begin{equation}
\label{Skl}\left\{ \varphi _1^M,\varphi _2^M\right\} _R=\left\langle R\
\nabla _{\varphi _1},\ \ \nabla _{\varphi _2}\right\rangle -\left\langle
R(AdM\cdot \ \ \nabla _{\varphi _1}),AdM\cdot \ \ \nabla _{\varphi
_2}\right\rangle .
\end{equation}
The right hand side in (13) is the{\em \ Sklyanin bracket} on $G$
determined
by $R.$ Thus we get the following well known result.

\begin{theorem}
Assume that $R=-R^{*}.\ $Equip $G$ with the Sklyanin bracket determined
by $R$. Then the monodromy map preserves Poisson brackets.
\end{theorem}

\section{Nonultralocal Case}

Let us now pass to the study of the general case when $R\neq -R^{*}.$
Lemma 1 shows that in general the Frechet derivative of a functional
$\varphi ^M$
is a discontinuous functions. Thus the problem is to extend the bilinear
form associated with the unbounded operator ${\cal H_{{\it R}}}$ to
functions with a jump. The problem of this kind is well known in
elementary
quantum mechanics (zero range potentials). The difference is that
in quantum
mechanics the free parameters of the model are the boundary conditions
imposed on the wave function. In our case, by contrast, the boundary
conditions are fixed in advance (as in Lemma 1 (iii)). The 'physical'
freedom consists in adding to the bilinear form of ${\cal H_{{\it R}}}$ an
interaction term which is sensitive to the jump of the Frechet
derivative at
$0\equiv 2\pi $. The resulting bilinear form usually has a lower symmetry
than the formal differential expression (12); hence one can speak of the
'spontaneous symmetry breaking'.

In order to describe the boundary bilinear form let us introduce the
following definition.

Let ${\bf g\oplus g}$ be the direct sum of two copies of ${\bf g}$. We
shall
equip ${\bf g\oplus g\ \ }$with the inner product
\begin{equation}
\label{Inprod}\left\langle \left\langle \left( X_1,Y_1\right) ,\left(
X_2,Y_2\right) \right\rangle \right\rangle =\left\langle
X_1,X_2\right\rangle -\left\langle Y_1,Y_2\right\rangle .
\end{equation}
Let us define a mapping
\begin{equation}
\label{d}\partial :C^\infty (\left[ 0,2\pi \right] ;{\bf g)\longrightarrow
g\oplus g:}X\longmapsto (X(0),X(2\pi )).
\end{equation}
Observe that for $X={\rm grad}\varphi ^M$ we have%
$$
\partial X=\left( X(0),AdL\cdot X(0)\right) ,
$$
and hence $\partial X$ is isotropic with respect to the inner product (14).

Let us define the bracket of two (in general, non smooth) functionals $%
\varphi _1,\varphi _2$ by the formula
\begin{equation}
\label{Br}\left\{ \varphi _1,\varphi _2\right\} =\frac 12\int(\left\langle
{\cal H}X_1,X_2\right\rangle -\left\langle {\cal H}X_2,X_1\right\rangle
)dx+\left\langle \left\langle B\partial X_1,X_2\right\rangle \right\rangle ,
\end{equation}
where $B\in {\rm End(}{\bf g\oplus g).}$ The bracket (16) is skew if
$B$ is
skew with respect to the inner product (14) on ${\bf g\oplus g.}$ Let us
discuss the conditions to be imposed on $B$ so as to make (16) a
{\em bona fide} Poisson bracket.

The necessary conditions on $B$ are as follows:

{\em (i) The boundary form should vanish on the diagonal subalgebra}
 {\bf g}$^\delta \subset $ ${\bf g\oplus g}${\em ${\bf .}$}${\bf (}$
  Indeed, if $\partial X_\varphi \in $ {\bf g$^\delta $},
  the functional $\varphi $ is
smooth.)

{\em (ii) ('Weak field approximation') The Poisson bracket vanishes
identically for }$M=1$; {\em its linearization at the unit
element} $1\in G$ {\em should coincide with the Lie--Poisson bracket
of the Lie algebra} ${\bf g_{{\it R}}.}$

The second condition is less obvious and deserves some comment. Observe
first of all that if the potential $L=0,$ the monodromy is equal to
identity. In this case the gradient of any functional $\varphi ^M$
is smooth
(cf. the boundary condition (11)), and the Poisson bracket is identically
zero (as before, the integrand in the formula for the Poisson bracket is a
total derivative of a periodic function).\ Assume that there exists a
Poisson bracket on $G$ which is compatible with the monodromy map; in that
case this bracket should, for consistency, also vanish at the unit element
of $G$. Moreover, if the potential $L$ in the auxiliary linear
equation (2)
is close to zero, we may find the monodromy perturbatively, and this
allows
to compute the linearization of the bracket at $M=1$. The result is quite
obvious: the linearized bracket coincides with the Lie--Poisson bracket of
the Lie algebra ${\bf g}_R$. If the r-matrix $R$ is skew, the Poisson
bracket on $G$ satisfying this condition is obviously the Sklyanin
bracket.
It is natural to impose this condition in general case as well; the
existence of such a Poisson bracket is of course nontrivial.

\begin{proposition}
An operator $B$ satisfying the above condition has the following form in
block notation
\begin{equation}
\label{Bform}B=\left|
\begin{array}{cc}
\alpha  & \alpha +s \\
-\alpha +s & -\alpha
\end{array}
\right| ,
\end{equation}

where $s=\frac 12(r+r^{*}),\alpha \in {\rm End(}{\bf g)}$ is a skew
symmetric operator.
\end{proposition}

We see in particular that the boundary form is needed in order to
correctly
reproduce the linearized bracket. Operator $\alpha $ is a free parameter
which characterizes the interaction term in our bilinear form; further
restrictions on $\alpha $ are imposed by the Jacobi identity.

\begin{proposition}
Let the bracket $\left\{ ,\right\} $ be defined by formula (\ref{Br}) with $%
{\cal H}$ given by (\ref{Op}) and the boundary form $B$ chosen as above.
Then the bracket of two functionals $\varphi _1^M,\varphi _2^M$ (where as
usual $\varphi _1^M,\varphi _2^M$ are smooth functions of the monodromy)
is
given by
\begin{equation}
\label{regbr}\left\{ \varphi _1^M,\varphi _2^M\right\} =\left\langle
\left\langle {\cal R}\partial X_1,\partial X_2\right\rangle \right\rangle ,
\end{equation}

where
\begin{equation}
\label{RR}{\cal R}=\left|
\begin{array}{cc}
a+\alpha  & \alpha +s \\
-\alpha +s & a-\alpha
\end{array}
\right| ,a=\frac 12\left( R-R^{*}\right) ,s=\frac 12\left( R+R^{*}\right) ,
\end{equation}

and $X\ _i={\rm grad}\varphi _i^M.$
\end{proposition}

Let ${\cal R}\in {\rm End}\left( {\bf g}\oplus {\bf g}\right) .$ define
the
bilinear map%
$$
\ \left[ \left[ {\cal R,R}\right] \right] :\wedge ^2\left( {\bf g}\oplus
{\bf g}\right) \longrightarrow {\bf g}\oplus {\bf g}
$$
by
$$
\left[ \left[ {\cal R,R}\right] \right] \left( X,Y\right) =%
\left[ {\cal R}X,%
{\cal R}Y\right] -{\cal R}(\left[ {\cal R}X,Y\right] +\left[ X,{\cal R}%
Y\right] ).
$$
The inner product on ${\bf g\ }$allows to identify $\left%
[ \left[ {\cal R,R}%
\right] \right] $ with an element of $\otimes ^3\left%
( {\bf g}\oplus {\bf g}%
\right) ;$ it is easy to see that if $R{\cal \ }$is skew, then actually $%
\left[ \left[ {\cal R,R}\right] \right] \in \wedge ^3\left( {\bf g}\oplus
{\bf g}\right) .$

\begin{proposition}
The bracket (\ref{regbr}) satisfies the Jacobi identity if and only if the
element
$\left[ \left[ {\cal R,R}\right] \right] \in \wedge ^3\left( {\bf g}%
\oplus {\bf g}\right) $ is $ad\left( {\bf g}\oplus {\bf g}\right) -$%
invariant.
\end{proposition}

As usual (cf. \cite{S}), it is convenient to replace this necessary and
sufficient condition with the following sufficient one.

\begin{proposition}
Assume that ${\cal R}$ satisfies the modified classical Yang--Baxter
equation
\begin{equation}
\label{CYB}\left[ {\cal R}X,{\cal R}Y\right] -{\cal R}(\left[ {\cal R}%
X,Y\right] +\left[ X,{\cal R}Y\right] )+\left[ X,Y\right] =0
\end{equation}
for any $X,Y\in {\bf g}\oplus {\bf g.\ }$Then the bracket (\ref{regbr})
satisfies the Jacobi identity.
\end{proposition}

It is useful to write down equation (\ref{CYB}) in terms of the matrix
coefficients of%
$$
{\cal R}=\left|
\begin{array}{cc}
A & B \\
B^{*} & D
\end{array}
\right| .
$$

\begin{proposition}
(i) Equation (\ref{CYB}) is equivalent to the following relations
\begin{equation}
\label{ABD}
\begin{array}[t]{lll}
\left[ Au,Av\right]  & = & A\left( \left[ Au,v\right] +\left[ u,Av\right]
\right) -\left[ u,v\right] , \\
\left[ Du,Dv\right]  & = & D\left( \left[ Du,v\right] +\left[ u,Dv\right]
\right) -\left[ u,v\right] , \\
\left[ Bu,Bv\right]  & = & B\left( \left[ Du,v\right] +\left[ u,Dv\right]
\right) , \\
\left[ B^{*}u,B^{*}v\right]  & = & B^{*}\left( \left[ Au,v\right] +\left[
u,Av\right] \right)
\end{array}
\end{equation}
for any $u,v\in {\bf g.}$

(ii) If relations (\ref{ABD}) are satisfied and moreover $A+B=B^{*}+D,$
then
$r=A+B$ satisfies the modified classical Yang--Baxter equation.
\end{proposition}

Relations (\ref{ABD}) on the matrix coefficients of $R{\cal \ }$ were
considered in \cite{Maillet}, \cite{Parm}; however, it passed unnoticed
that
they are equivalent to the modified CYBE for the square of ${\bf g}$ which
reduces the classification of solutions to a standard problem.

Let us recall the fundamental classification theorem of Belavin and
Drinfeld
\cite{BelDr}.

\begin{theorem}
Let ${\bf g}$ be an affine Lie algebra, ${\bf h}\subset {\bf g}$ its Cartan
subalgebra, $P\subset {\bf h}^{* }$the set of its simple roots. (i)To
each solution of equation (\ref{CYB}) on ${\bf g}$ one can assign a
triple
$(\Gamma _1,\Gamma _2,\tau )$,
where $\Gamma _1,\Gamma _2\subset P$ and $\tau$
is an isometry $\Gamma _1\longrightarrow \Gamma _2$ such that
$\tau^k\alpha \notin \Gamma _1$ for any $\alpha \in \Gamma _1$ and for
sufficiently large $k$ (expression $\tau ^k\alpha $ makes sense if
$\tau\alpha ,\tau ^2\alpha ,...,\tau ^{k-1}\alpha \in \Gamma _1%
\cap \Gamma _2).$
(ii) For each triple $(\Gamma _1,\Gamma _2,\tau )$ the solutions are
parametrized by tensors $r\in {\bf h}\otimes {\bf h}$ such that%
$$
r_{12}+r_{21}=t_0
$$
is a Casimir element in ${\bf h}\otimes {\bf h}$ and for each $\alpha \in
\Gamma _1$%
$$
(\tau \alpha \otimes id+id\otimes \alpha )r=0.
$$
\end{theorem}

The system of simple roots of ${\bf g}\oplus {\bf g}$ is the union of two
copies of $P$ (we shall denote the second copy by $\tilde P$). Let us give
three important examples of r-matrices on ${\bf g}\oplus {\bf g\ }$
satisfying the additional condition $A+B=B^{*}+D.$

(1) $\Gamma _1=\Gamma _2$ =$\ \emptyset ;$ in this case R=$\left|
\begin{array}{cc}
r & 0 \\
0 & r
\end{array}
\right| ,$ where $r$ is the standard 'trigonometrical' r-matrix associated
with ${\bf g.}$

(2) Let $\Gamma _1=P,\Gamma _2=\tilde P\;\;$and let $\tau $ be a natural
isometry $P\longrightarrow \tilde P$. The corresponding r-matrix has the
form
\begin{equation}
\label{Doub}R=\left|
\begin{array}{cc}
r & r_{+} \\
r_{-} & -r
\end{array}
\right|
\end{equation}
where $r$ is the same as above and $r_{\pm }=(r\pm id).$ The r-matrix
(\ref{Doub}) is the canonical r-matrix of the double of the Lie
bialgebra $({\bf g},{\bf g}_r)$.

(3) Let ${\bf g}={\cal L}({\bf a})$ be the loop algebra of a
semisimple Lie
algebra ${\bf a};$ let $\alpha $ be the root of ${\bf g}$ which
corresponds
to the additional vertex of the extended Dynkin diagram of ${\bf a}$.
Put $%
\Gamma _1=P\setminus \left\{ \alpha \right\} ,\Gamma _2$ $=\tilde P\setminus
\left\{ \tilde \alpha \right\} \ $ $\;$and let $\tau $ be a natural
isometry
$\Gamma _1\longrightarrow \Gamma _2.$ Let us denote by $r^0$ the standard
r-matrix on ${\bf a}\subset {\cal L}({\bf a})$. Then
\begin{equation}
R=\left|
\begin{array}{cc}
r+r_{\ }^0 & r_{+}^0 \\
r_{-}^0 & r-r^0
\end{array}
\right|
\end{equation}
Observe that in case (1) the bracket (\ref{regbr}) is the standard
Sklyanin
bracket; in case (2) it is the dual bracket on $G$ (cf. \cite{S2} and the
discussion in the next Section   below); case (3) is an interpolation
between the first two. The symmetric part of the linearized bracket is
zero
in case (1), in case (2) we have $s=id$, and in case (3) we have $s=P^0$,
where $P^0$ is the projection operator onto the subalgebra of constant
loops
${\bf a}\subset {\cal L}({\bf a}).$

\section{Conclusion. A few Words on Symmetry Breaking}

It is probably worth saying a few words on the resulting breakdown of
global
gauge symmetry and 'spontaneous quantization' of the global gauge group.
Let
${\bf G}=C^\infty (S^1,G)$ be the loop group of $G.$ We may identify $G$
with the subgroup of constant loops. Let ${\bf G}_0$ be the subgroup of $%
{\bf G}$ consisting of  loops satisfying $g(0)=e.$ Clearly, ${\bf G}_0$ is
normal in ${\bf G}$ and ${\bf G}_0{\bf \ /G\ =}G.$ The group ${\bf G}$
acts
on the space of the first order diffential operators (\ref{L}) by
conjugations, and this induces the gauge action of ${\bf G}$ on the phase
space ${\cal M}$ . Let $G\times G\longrightarrow G$ be the action of $G$
on
itself by conjugations. Clearly, we get a commutative diagram
$$
\begin{array}{cccccc}
{\bf \ } & \ {\bf G}\times {\cal M} & \  & \longrightarrow  & {\cal M} &  \\
\pi \times M & \downarrow  &  &  & \downarrow  & M \\
\  & \ G\times G & \ \  & \longrightarrow  & G &
\end{array}
$$
Suppose now that $G$ is a finite-dimensional simple Lie group and the
r-matrix is chosen as in Example 2 of the previous section, i.e., it is
given by (\ref{Doub}). The Poisson bracket for the monodromy matrices we
get
in this way is essentially that of the {\em dual} group $G^{*}$ which is
identified with $G$ via the canonical factorization map. Let us equip the
group $G$ of global gauge transformations with the standard Sklyanin
bracket. According to the well known results of the Poisson Lie groups
theory \cite{S2}, there is a canonical Poisson action $G\times
G^{*}\longrightarrow G^{*}$ called dressing transformations. As explained
in
\cite{S2}, if we identify $G^{*}$ with $G,$ dressing transformations
correspond to conjugations in $G;$ thus in order to maintain the gauge
covariance of our monodromy map (which should, we recall, be a morphism in
the category of Poisson manifolds) we must equip the global gauge group
with
a nontrivial Poisson bracket. By contrast, it is consistent to assume that
the subgroup ${\bf G}_0$ remains classical, i.e. carries a trivial Poisson
bracket. The implications for quantization are obvious: to preserve the
gauge covariance on the quantum level we have to assume that the global
gauge group becomes quantum (while the subgroup ${\bf G}_0$ remains
classical. If we replace functions on the circle with the more physical
case
of functions on the line, the same discussion will apply to the subgroup
of
rapidly decreasing gauge transformations.

In applications to completely integrable systems we usually need a Lax
operator with a spectral parameter, i.e. our Lie algebra {\bf g} is a loop
algebra in auxiliary parameter $\lambda .$ The model situation here is
illustrated by Example 3 of the previous section. It is natural to
demand in
this case the Poisson covariance of the monodromy map with repect to the
global gauge group consisting of functions which do not depend neither
on $x$
nor on $\lambda .$ This is again made consistent with the regularized
Poisson bracket for the monodromy provided that we equip the global gauge
group with the Sklyanin bracket which corresponds to the constant
r-matrix $%
r^0\ .$

Let us outline our conclusions. If the input r-matrix of the model is not
skew, the Poisson bracket for the monodromy requires regularization; the
regularized Poisson bracket is determined by a solution of the modified
classical Yang--Baxter equation on the square of the Lie algebra $g$. All
such solutions may be completely classified. Applications to concrete
examples (in particular, to lattice systems and difference Lax equations)
will be considered in a separate paper. ( A special case corresponding to
example (2) above was studied in \cite{AFSV},\cite{AFS}.)


\begin{thebibliography}{99}
\bibitem{FadTak}  Faddeev\ L.D., Takhtajan L.A.,{\em \ Hamiltonian methods
in the theory of solitons,} Springer, 1986.

\bibitem{BelDr}  Belavin A.A., Drinfeld V.G., {\em On solutions of the
classical Yang-Baxter equation for simple Lie algebras.} Funct.
Anal.\ Appl.
{\bf \ 16}, 1982, 159--182.

\bibitem{ReyS}  Reyman A.G., Semenov-Tian-Shansky M.A., {\em Current
algebras and nonlinear equations.} Sov. Math. Dokl. {\bf 21}, 1980,
630--634.

\bibitem{S}  Semenov-Tian-Shansky M.A., {\em What is a classical
r-matrix. }
Funct. Anal.\ Appl. {\bf \ 17,} 1983, 259--272.

\bibitem{S2}  Semenov-Tian-Shansky M.A.,{\it \ Dressing transformations
and
Poisson group actions, }Publ. Res. Inst. Math. Sci. Kyoto University, {\bf 21%
}, 1985, 1237--1260.

\bibitem{Alv}  Alvarez-Gaum\'e L., Gomez C., Sierra G., {\em Duality and
quantum groups. }Nucl.\ Phys. {\bf B330,} 1990, 347-398.

\bibitem{Bab}  Babelon O., {\em Extended conformal algebra and Yang-Baxter
equation.}Phys.\ Lett. {\bf B215,} 1988, 523-529.

\bibitem{Gaw}  Gawedzki K., {\em Classical origin of quantum group
symmetries.} Preprint I.H.E.S. P/90/92,\ 1990.

\bibitem{AFSV}  Alekseev\ A., Faddeev L.D., Semenov-Tian-Shansky M.A.,
Volkov A.,{\em \ The unravelling of the quantum group structure in the WZNW
theory}.\ Preprint CERN-TH -5981/91, 1991.

\bibitem{AFS}  Alekseev\ A., Faddeev L.D., Semenov-Tian-Shansky M.A.,
{\em %
Hidden quantum groups inside Kac-Moody algebras, }Commun.\ Math. Phys.
149,
1992, 335-345.

\bibitem{Maillet}  Freidel L., MailletJ.-M.,{\em \ Quadratic Poisson
brackets%
}. Preprint LPTHE-24/91, Paris 1991.

\bibitem{Parm}  Parmentier S., {\em Twisted affine Poisson structures,
decompositions of Lie algebras and the classical Yang-Baxter equation. }
Preprint. Max Planck Inst. f\"ur Mathematik, Bonn 1991.
\end{thebibliography}
\end{document}